%% file: TOP-11-017_temp.tex
\pdfoutput=1

\documentclass[11pt,twoside,a4paper,cmspaper,final,collab]{cms-tdr}
\def\svnVersion{235428}\def\cmsCernNo{2013-124}\def\cmsCernDate{\today}\def\cmsMessage{Published in the European Physical Journal C as \href{http://dx.doi.org/10.1140/epjc/s10052-014-2758-x}{\doi{10.1140/epjc/s10052-014-2758-x.}}}

\begin{document}\cmsNoteHeader{TOP-11-017}

\hyphenation{had-ron-i-za-tion}
\hyphenation{cal-or-i-me-ter}
\hyphenation{de-vices}

\RCS$Revision: 224371 $
\RCS$HeadURL: svn+ssh://svn.cern.ch/reps/tdr2/papers/TOP-11-017/trunk/TOP-11-017.tex $
\RCS$Id: TOP-11-017.tex 224371 2014-01-24 12:24:11Z alverson $
\newlength\cmsFigWidth
\ifthenelse{\boolean{cms@external}}{\setlength\cmsFigWidth{0.85\columnwidth}}{\setlength\cmsFigWidth{0.4\textwidth}}
\ifthenelse{\boolean{cms@external}}{\providecommand{\cmsLeft}{top}}{\providecommand{\cmsLeft}{left}}
\ifthenelse{\boolean{cms@external}}{\providecommand{\cmsRight}{bottom}}{\providecommand{\cmsRight}{right}}
\ifthenelse{\boolean{cms@external}}{\providecommand{\breakhere}{\linebreak[4]}}{\providecommand{\breakhere}{\relax}}
\cmsNoteHeader{TOP-11-017} 
\title{Measurement of the top-quark mass in all-jets $\ttbar$ events in pp collisions at $\sqrt{s}=7$\TeV}

\date{\today}

\abstract{
The mass of the top quark is measured using a sample of $\ttbar$ candidate events with at least six jets in the final state. The sample is selected from data collected with the CMS detector in pp collisions at $\sqrt{s}=7$\TeV in 2011 and corresponds to an integrated luminosity of 3.54\fbinv. The mass is reconstructed for each event employing a kinematic fit of the jets to a $\ttbar$ hypothesis. The top-quark mass is measured to be $173.49\pm0.69\stat\pm1.21\syst$\GeV. A combination with previously published measurements in other decay modes by CMS yields a mass of $173.54\pm0.33\stat\pm0.96\syst$\GeV.
}

\hypersetup{%
pdfauthor={CMS Collaboration},%
pdftitle={Measurement of the top-quark mass in all-jets t t-bar events in pp collisions at sqrt(s)=7 TeV},%
pdfsubject={CMS},%
pdfkeywords={CMS, physics, top}}

\maketitle 
\section{Introduction}
The mass of the top quark ($m_\cPqt$) is an essential parameter of the standard model.
Its measurement also provides an important benchmark for the performance and calibration of the Compact Muon Solenoid (CMS) detector~\cite{CMS:2008zzk} at the CERN Large Hadron Collider (LHC).
The top-quark mass has been determined with high precision at the Fermilab Tevatron~\cite{PhysRevD.86.092003} to be $m_\cPqt=173.18\pm0.94$\GeV.
Measurements have been carried out in several top-quark decay channels using different methods, with the most precise single measurement at the Tevatron being that performed by the CDF Collaboration~\cite{Aaltonen:2012zzz} in the lepton+jets final state using a template method yielding $m_\cPqt=172.85\pm1.11$\GeV.

In this article a measurement is presented  using a sample of \ttbar candidate events with six or more reconstructed jets in the final state. It represents the first mass measurement in the all-jets channel performed by the CMS Collaboration.
The all-jets decay mode has a larger signal yield than the dilepton and lepton+jets channels.
However, with only jets in the final state, this channel is dominated by a multijet background and this measurement requires dedicated triggers and tight selection criteria.
This measurement complements the latest measurements by the CMS Collaboration in the lepton+jets and dilepton channels that yield $m_\cPqt = 173.49 \pm 1.07$\GeV~\cite{Chatrchyan:2012cz} and $m_\cPqt = 172.5 \pm 1.5$\GeV~\cite{Chatrchyan:2012ea}, respectively.
The most precise measurement in the all-jets channel so far is by the CDF Collaboration yielding $m_\cPqt=172.5\pm2.0$\GeV~\cite{Aaltonen:2011em}.

The event selection is very similar to the one used for the CMS \ttbar cross section measurement in the same final state, requiring at least six jets~\cite{Chatrchyan:2013ual}. Analogously to  the CMS measurement of the top-quark mass in the lepton+jets channel~\cite{Chatrchyan:2012cz}, the analysis employs a kinematic fit of the decay products to a \ttbar hypothesis and likelihood functions for each event (``ideograms'') that depend on the top-quark mass only or on both the top-quark mass and the jet energy scale.

\section{CMS Detector}
\label{sec:detector}
The central feature of the CMS apparatus is a superconducting solenoid, of 6\unit{m} internal diameter, providing a field of 3.8\unit{T}.
The bore of the solenoid is equipped with various particle detection systems.
CMS uses a right-handed coordinate system, with the origin at the nominal interaction point, the $x$ axis pointing to the center of the LHC ring, the $y$ axis pointing up (perpendicular to the plane of the LHC ring), and the $z$ axis along the counterclockwise-beam direction.
The polar angle, $\theta$, is measured from the positive $z$ axis and the azimuthal angle, $\phi$, is measured in the $x$-$y$ plane in radians.

{\tolerance=1000
Charged-particle trajectories are measured with silicon pixel and strip trackers, covering the pseudorapidity range $\abs{\eta} <2.5$, where $\eta \equiv -\ln[\tan(\theta/2)]$.
A lead-tungstate crystal electromagnetic calorimeter (ECAL) and a brass/scintillator hadron calorimeter (HCAL) surround the tracking volume.
The HCAL, when combined with the ECAL, measures jets with a resolution \breakhere$\Delta E/E \approx 100\% / \sqrt{\smash[b]{E\,[\GeVns]}} \oplus 5\%$.
In addition to the barrel and endcap detectors, CMS has extensive forward calorimetry that extends the coverage to $\abs{\eta} < 5$.
Muons are measured up to $\abs{\eta}<2.4$ using gas-ionization detectors embedded in the steel flux-return yoke outside the solenoid.
A two-level trigger system selects the final states pertinent to this analysis.
A detailed description of the CMS detector is available elsewhere~\cite{CMS:2008zzk}.
\par}

\section{Data Samples and Event Selection}
\label{sec:eventselection}
The analyzed data sample has been collected in 2011 in pp collisions at $\sqrt{s}=7$\TeV using two different multijet triggers and corresponds to an  integrated luminosity of $3.54 \pm 0.08\fbinv$~\cite{CMS-PAS-SMP-12-008}. The first trigger requires the presence of at least four jets built only from the energies deposited in the calorimeters with transverse momenta $\pt \ge 50$\GeV and the presence of a fifth calorimeter jet with $\pt \ge 40$\GeV. An additional requirement of a sixth calorimeter jet with $\pt \ge 30$\GeV was added during the data taking and this second trigger collected 3.19\fbinv of data.

Our procedure uses simulated events to estimate the composition of the data sample, to determine and calibrate the ideograms, and to evaluate the systematic uncertainties. The \ttbar signal events have been generated for nine different top-quark mass values ranging from 161.5 to 184.5\GeV with the \MADGRAPH~5.1.1.0 matrix element generator~\cite{Alwall:2011uj}, \PYTHIA~6.424 parton showering~\cite{Sjostrand:2006za} using the Z2 tune~\cite{Chatrchyan:2011id}, and a full \GEANTfour~\cite{Agostinelli:2002hh} simulation of the CMS detector.
The matching between the matrix elements (ME) and the parton shower evolution (PS) is done by applying the MLM prescription described in Ref.~\cite{Mangano:2006rw}.
The simulation includes the effects of additional overlapping minimum-bias events (pileup) so that the distribution of the number of proton interactions per bunch crossing matches the corresponding distribution in data. Furthermore, the jet energy resolution in simulation has been scaled to match the resolution observed in data~\cite{Chatrchyan:2011ds}.

Jets are formed by clustering the particles reconstructed by a particle-flow algorithm~\cite{CMS-PAS-PFT-10-002} using the anti-\kt algorithm~\cite{Cacciari:2008gp,Cacciari:2011ma} with a radius parameter of 0.5.
The particle-flow technique combines information from all subdetectors to reconstruct individual particles including muons, electrons, photons, charged hadrons, and neutral hadrons. It typically improves the jet energy resolution to 15\% at 10\GeV, 8\% at 100\GeV, and 4\% at 1\TeV.
An additional advantage of this technique is that it facilitates pileup removal by discarding charged particles associated with vertices other than the primary and secondary vertices from the primary collision.
Jet energy corrections are applied to all the jets in data and simulation~\cite{Chatrchyan:2011ds}. These corrections are derived from simulation and are defined as a function of the transverse momentum density of an event~\cite{Cacciari:2007fd,Cacciari:2008gn,Cacciari:2011ma} as well as of the \pt and $\eta$ of the reconstructed jet. By these means a uniform energy response at the particle level with low pileup dependence is obtained.
A residual correction, measured from the momentum balance of dijet and $\gamma$+jet/Z+jet events, is applied to the jets in data.
To reduce the contamination by false jets from detector noise or by electrons reconstructed as jets, the fractions of the jet energy from photons, electrons, and neutral hadrons are required to be below 99\%, and the fraction of the jet energy from charged hadrons is required to be greater than zero.

Since hadronically decaying top-quark pairs lead to six quarks in the final state, events are selected with at least four jets with $\pt > 60$\GeV, a fifth jet with $\pt > 50$\GeV, and a sixth jet with $\pt > 40$\GeV.  Additional jets are considered only if they have  $\pt > 30$\GeV. All jets are required to be within pseudorapidity $\abs{\eta}$ of 2.4, where the tracker acceptance ends. The Combined Secondary Vertex tagger with the Tight working point (CSVT)~\cite{Chatrchyan:2012jua} is used to tag jets originating from bottom quarks. The CSVT working point corresponds to an efficiency of approximately 60\%, while the misidentification probability for jets originating from light quarks (uds) and gluons is only 0.1\%. We require at least two b-tagged jets. After these initial event selection criteria, 26\,304 candidate events are selected in the data.

\section{Kinematic Fit}
\label{sec:kinfit}
For the final selection of candidate \ttbar events, a kinematic least-squares fit~\cite{kinfitCMS} is applied.
It exploits the characteristic topology of \ttbar events: two W~bosons that can be reconstructed from the untagged jets and two top quarks that can be reconstructed from the W~bosons and the b-tagged jets.
The reconstructed masses of the two top quarks are constrained to be equal. In addition, the mass of both W bosons in the event is constrained to 80.4\GeV~\cite{pdg} in the fit leading to $n_{\text{dof}}=3$ degrees of freedom.
Gaussian resolutions are used for the jet energies in the kinematic fit.
They are separately determined for jets originating from light quarks and bottom quarks as functions of \pt and $\eta$ using simulated \ttbar events.

To find the correct combination of jets, the fit procedure is repeated for every experimentally distinguishable jet permutation.
This is done using all (six or more) jets that pass the selection. In the data, 8810 events have exactly seven selected jets, 3259 events have eight jets, and 1183 events have nine or more jets.
All b-tagged jets are taken as bottom-quark candidates, the untagged jets serve as light-quark candidates.
If the fit converges for more than one of the possible jet permutations, the one with the smallest fit $\chi^2$ is chosen.
After the kinematic fit, all events with a goodness-of-fit probability of $P_\text{gof} = P\left(\chi^2,n_{\text{dof}}=3\right) > 0.09$ are accepted.

{
To further reduce the multijet background with $\bbbar$ production, an additional criterion on the separation of the two bottom-quark candidates, $\Delta R_{\bbbar} = \breakhere\sqrt{\smash[b]{(\Delta \phi_{\bbbar})^2+(\Delta \eta_{\bbbar})^2}}> 1.5$, is imposed.
The number of events in data passing each selection step, the expected fraction of signal events in the data sample assuming a \ttbar cross section of 163\unit{pb}~\cite{Kidonakis:2010dk}, and the selection efficiency for signal are given in Table~\ref{tab:cutFlow}.
\par}

\begin{table}[htb]
\centering{
\topcaption[Cutflow]{Number of events, the predicted signal fraction in the data sample, and the selection efficiency for signal after each selection step. The predicted signal fraction is derived from simulation assuming a \ttbar cross section of 163\unit{pb}~\cite{Kidonakis:2010dk} and a top-quark mass of 172.5\GeV.}
\begin{tabular}{l r r r}
 \hline
  Selection step      & Events  & Sig. frac. & Sel. eff.\\
  &&& for signal\\
 \hline\hline
  At least 6 jets                     & 786\,741 & 3\% & 3.48\%  \\
  At least two b tags                 &  26\,304 &  17\% & 0.91\% \\
  $P_\text{gof} > 0.09$                &   3\,691 &  39\% & 0.30\% \\
  $\Delta R_{\bbbar}>1.5$ &   2\,418 &  51\% & 0.25\% \\
 \hline
\end{tabular}
\label{tab:cutFlow}
}
\end{table}

To extract the mass, the events are weighted by their goodness-of-fit probabilities increasing the fraction of \ttbar events to $54$\% and improving the resolution of the fitted top-quark mass.
We classify the \ttbar events based on the jet-parton associations in simulation. Partons are matched to a jet if they are separated by less than 0.3 in $\eta$-$\phi$ space.
Three different categories are distinguished in the following way: correct permutations $cp$ (27.9\%), wrong permutations $wp$ (22.6\%) where at least one jet is not associated to the correct parton from the \ttbar decay, and unmatched permutations $un$ (49.4\%).
The last case contains events in which at least one quark from the \ttbar decay cannot be matched unambiguously to a selected jet.
For correct permutations, the kinematic fit and the weighting procedure improve the resolution of the fitted top-quark masses from 13.6\GeV to 7.9\GeV.
Furthermore, the requirement on the goodness-of-fit probability removes 76\% of the signal events classified as unmatched permutations enhancing the fraction of correct permutations from 10\% to 27.9\%.

\section{Background Modeling}
\label{sec:signalbackground}

The multijet background is estimated using an event mixing technique.
All events after the b-tagging selection are taken as input.
The jets are mixed between the different events based on their position in a \pt-ordered list in the event in which they were recorded; every jet in the events in the multijet background model originates from a different event in the data, with the \pt-ordered position preserved. No duplicate jets, in terms of their \pt-ordering, are allowed.
In addition, it is required that at least two b-tagged jets are found in every new event. The kinematic fit to  a \ttbar hypothesis is performed on each mixed event and the same $P_\text{gof}$ and $\Delta R_{\bbbar}$ selection is applied.
This procedure was validated on particle-level jets using \bbbar events generated with \PYTHIA. The distributions of the fitted top-quark mass $m_\cPqt^\text{fit}$ and the mean of the two reconstructed \PW-boson masses agree well between the generated \bbbar events and the modeled events from event mixing on the same sample.

As can be seen in Table~\ref{tab:cutFlow}, the input sample has an expected fraction of 17\% \ttbar events. The impact of this contamination on the background prediction is evaluated with simulated \ttbar events and its minor effect on the background modeling is treated as a systematic uncertainty.

We normalize the simulated \ttbar sample and the background prediction to data with an expected signal fraction $f_{\text{sig}}$ from simulation.
This signal fraction $f_{\text{sig}}$ depends on the \ttbar cross section and the selection efficiency for \ttbar events for different top-quark masses.
It varies between 50\% and 55\% for top-quark masses within three standard deviations of the Tevatron average top-quark mass~\cite{PhysRevD.86.092003} for three different predictions of the \ttbar cross section~\cite{Ahrens:2010zv,Aliev:2010zk,Kidonakis:2010dk}.
Adding to this the uncertainty in the luminosity and the systematic uncertainties in the selection efficiency~\cite{Chatrchyan:2013ual}, we assume $f_{\text{sig}} = (54\pm 4\,(\text{th.}) \pm 1\lum\pm 10\syst)$\% for this analysis.

\begin{figure*}[htbp]
\centering
\includegraphics[width=0.48\textwidth]{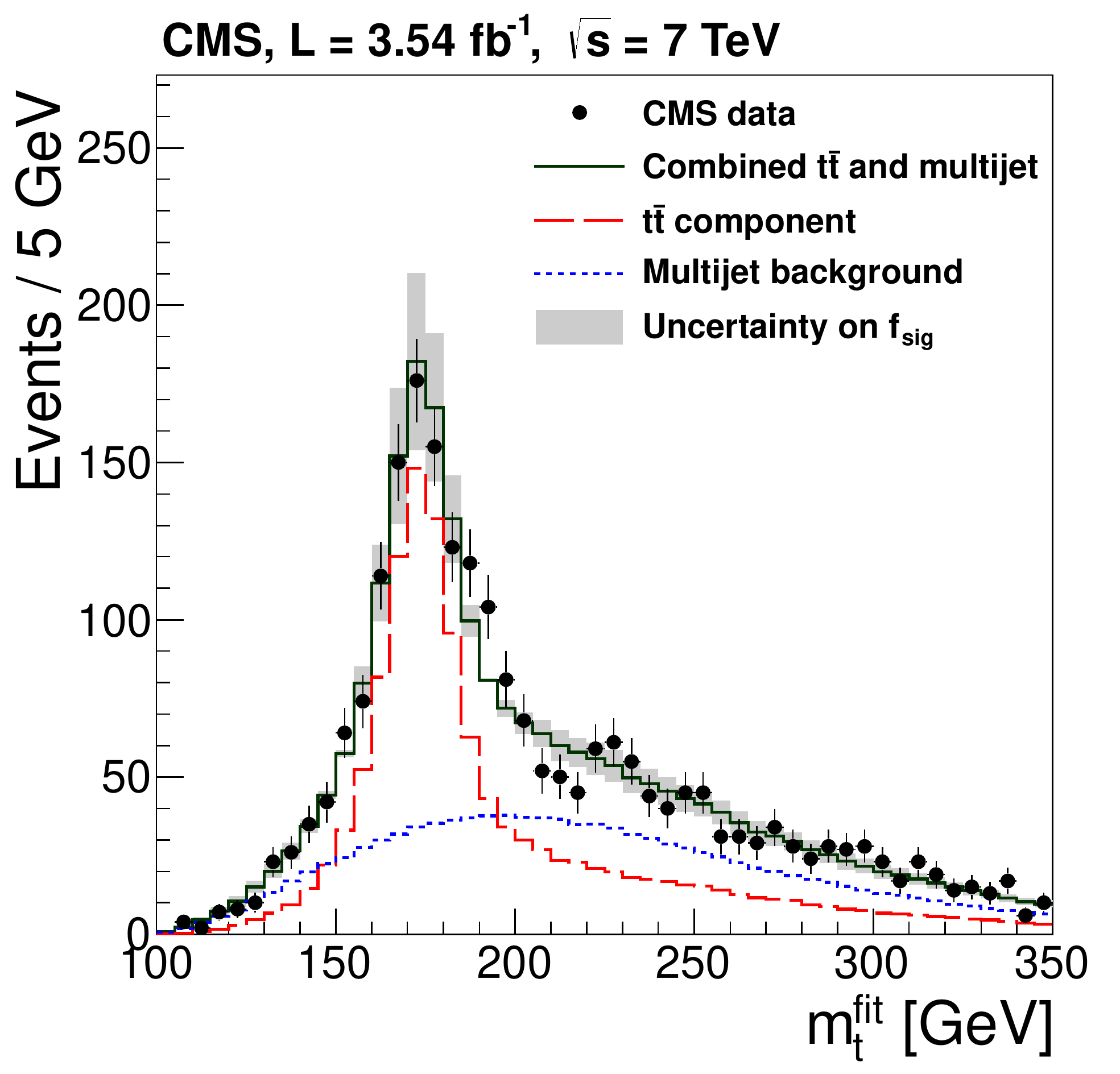}
\includegraphics[width=0.48\textwidth]{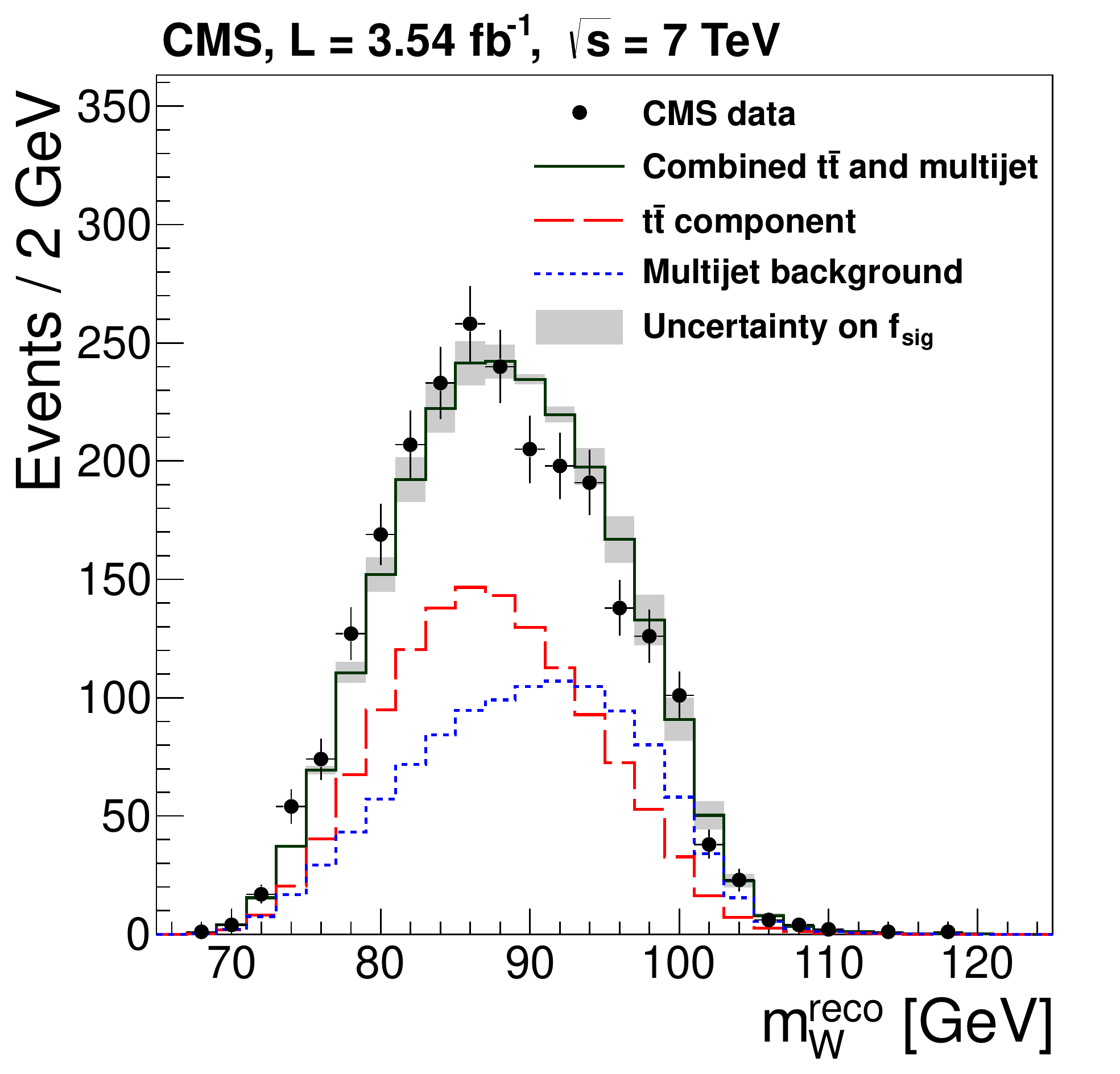}
\includegraphics[width=0.48\textwidth]{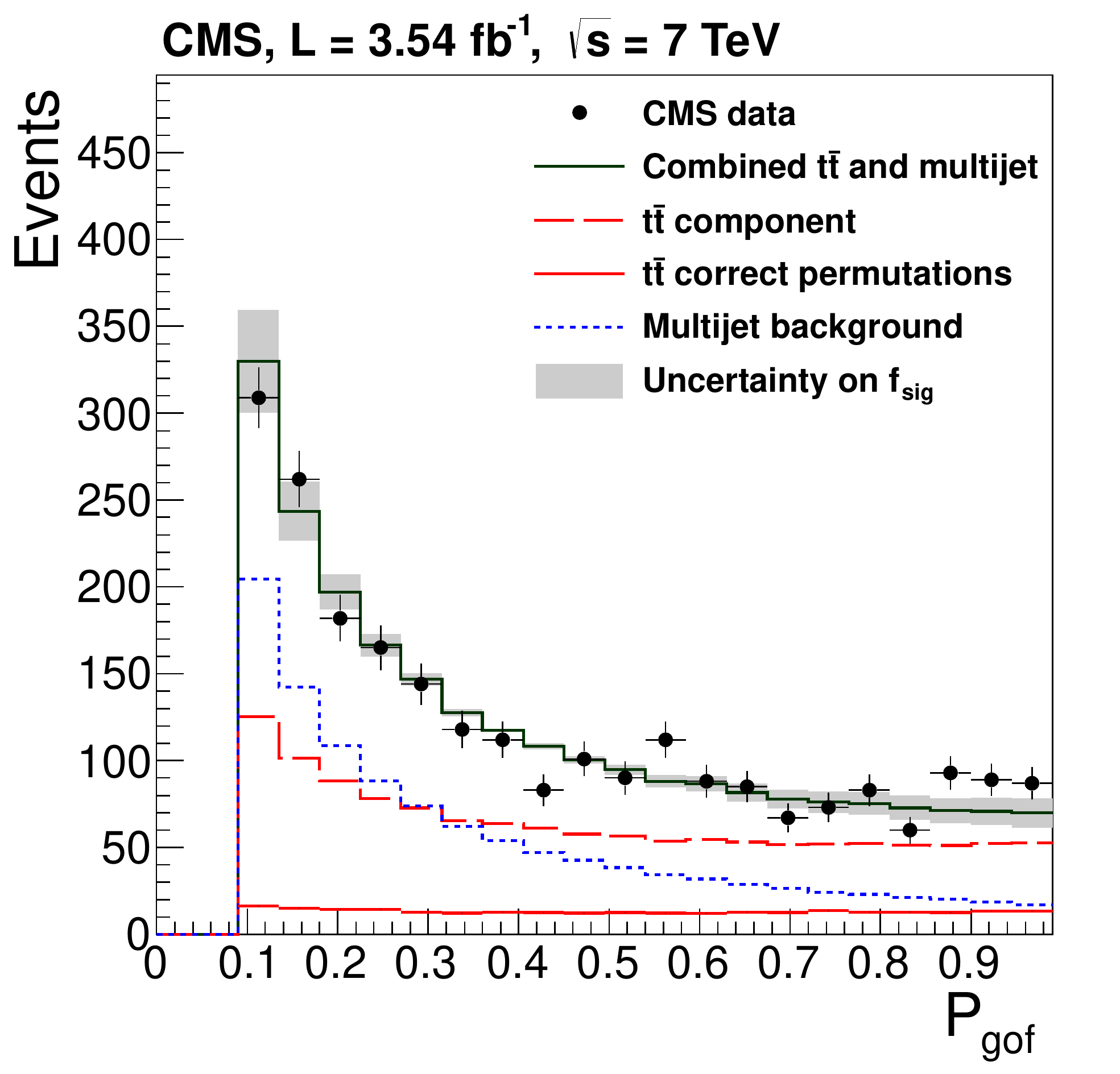}
\includegraphics[width=0.48\textwidth]{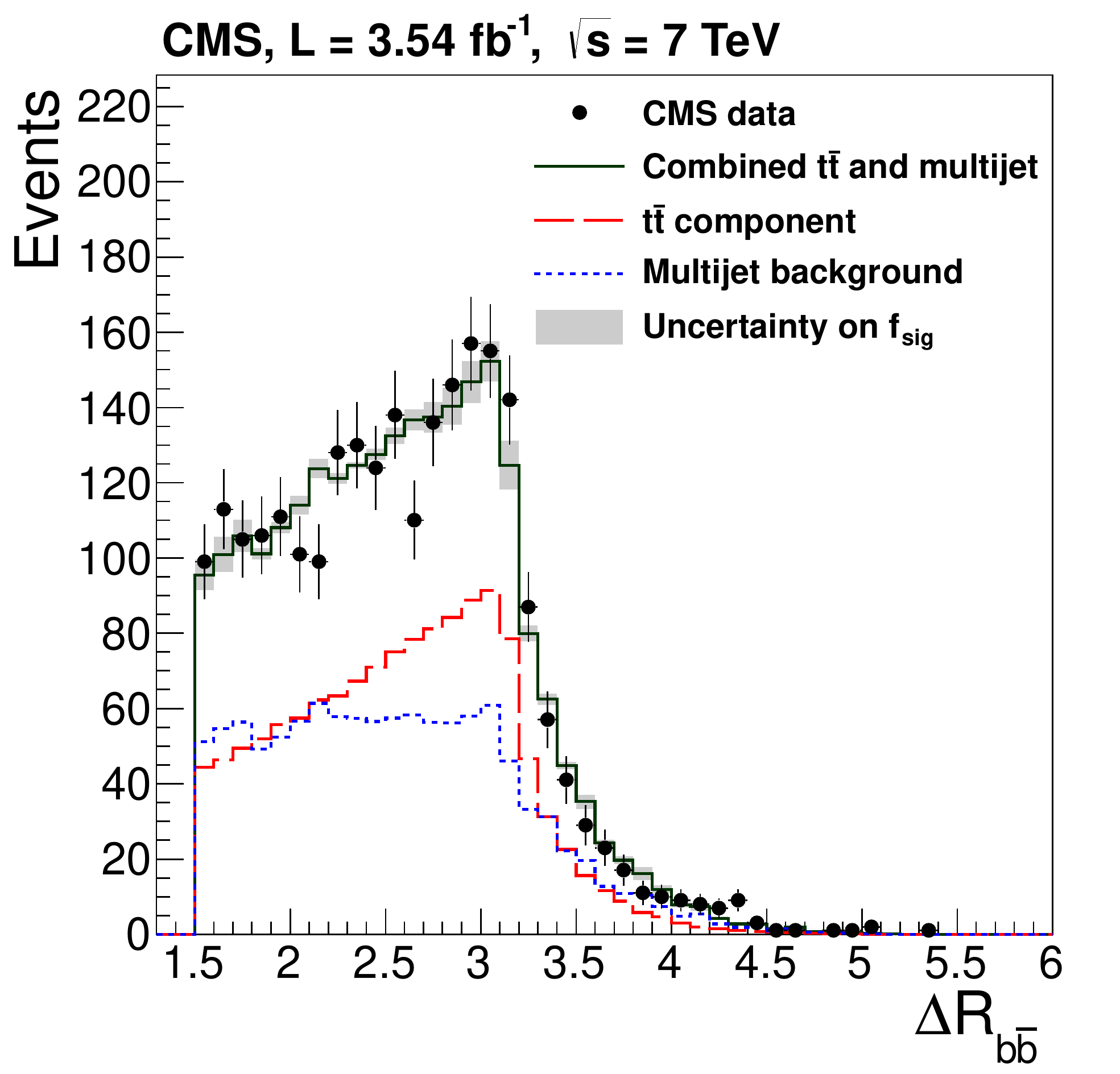}

\caption{\label{fig:controlplot-obs} (upper left) Reconstructed top-quark mass from the kinematic fit, (upper right) average reconstructed W-boson mass, (lower left) goodness-of-fit probability, and (lower right) the separation of the two b-tagged jets after all selection steps.
The simulated \ttbar signal and the background from event mixing are normalized to data.
The band indicates the correlated uncertainty from the signal fraction $f_\text{sig}$.
The top-quark mass used in the simulation is 172.5\GeV and the nominal jet energy scale is applied.}
\end{figure*}

Figure~\ref{fig:controlplot-obs} compares data and the expectation from simulation and background for the fitted top-quark mass $m_\cPqt^\text{fit}$, the mean of the two reconstructed W-boson masses per event $m_{\PW}^\text{reco}$, the goodness-of-fit probability $P_\text{gof}$ , and the distance between the two b-tagged jets $\Delta R_{\bbbar}$.
Overall, the agreement is good within the uncertainties.

\section{Ideogram Method}
\label{sec:ideogram}

Since the jet energy scale (JES) is the leading systematic uncertainty in previous top-quark mass measurements, we construct a likelihood function that allows the determination of the JES and the top-quark mass simultaneously by a joint fit to all selected events in data.
The JES is estimated from the invariant masses of the jets associated with the W bosons exploiting the precise knowledge of the W-boson mass from previous measurements~\cite{pdg}. Based on this likelihood function, we perform two different estimations of the top-quark mass: one with a fixed JES (henceforth ``1D analysis'') and a second with a simultaneous estimation of the JES (henceforth ``2D analysis'').
The 2D analysis is similar to the measurements of the top-quark mass in the all-jets channel by the CDF Collaboration~\cite{Aaltonen:2011em} and in lepton+jets final states by the CMS Collaboration~\cite{Chatrchyan:2012cz}.

The observable used for measuring $m_\cPqt$ is the top-quark mass $m_\cPqt^\text{fit}$ obtained from the fitted four-momenta of the jets after the kinematic fit.
We take the mean of the two reconstructed W-boson masses before they are constrained by the kinematic fit $m_\PW^\text{reco}$ as an estimator for measuring in situ an additional global JES beyond that of the standard CMS jet energy corrections.
The likelihood calculation in the ideogram method~\cite{Abdallah:2008xh,Aaltonen:2006xc,Abazov:2007rk} is done by evaluation of analytic expressions for the probability densities. These expressions are derived and calibrated using simulated events and the modeled background from event mixing.

A likelihood to estimate the top-quark mass and JES given the observation of a data sample can be defined as:

\ifthenelse{\boolean{cms@external}}{
\begin{equation}
\begin{aligned}
\mathcal{L}&\left(m_\cPqt,\mathrm{JES}|\text{sample}\right)  \propto  P\left(\text{sample}|m_\cPqt,\mathrm{JES}\right)\\
 &=  \prod_\text{events}P\left(m_\cPqt^\text{fit},m_\PW^\text{reco}|m_\cPqt,\mathrm{JES}\right)^{w_\text{event}}.\label{eq:1}\\
\end{aligned}
\end{equation}
}{
\begin{equation}
\mathcal{L}\left(m_\cPqt,\mathrm{JES}|\text{sample}\right)  \propto  P\left(\text{sample}|m_\cPqt,\mathrm{JES}\right)
 =  \prod_\text{events}P\left(m_\cPqt^\text{fit},m_\PW^\text{reco}|m_\cPqt,\mathrm{JES}\right)^{w_\text{event}}.\label{eq:1}\\
\end{equation}
}
The event weight $w_\text{event} \propto P_\text{gof}$ is introduced in order to lower the impact of unmatched and background events. The sum of all event weights is normalized to the number of events.

Due to the mass constraint on the W boson in the fit, the correlation coefficient between $m_\cPqt^\text{fit}$ and $m_\PW^\text{reco}$ is only $-0.08$ for correct permutations in simulation. Hence, we treat $m_\cPqt^\text{fit}$ and $m_\PW^\text{reco}$ as uncorrelated and the probability $P\left(m_{\cPqt}^\text{fit},m_\PW^\text{reco}|m_\cPqt,\mathrm{JES}\right)$ from  Eq.~(\ref{eq:1}) is factorized into
\ifthenelse{\boolean{cms@external}}{
\begin{align*}
P&\Big(m_\cPqt^\text{fit},m_\PW^\text{reco}|m_\cPqt,\mathrm{JES}\Big)  = {} \\
&f_\text{sig} \cdot P_\text{sig}\Big(m_\cPqt^\text{fit},m_\PW^\text{reco}|m_\cPqt,\mathrm{JES}\Big)\\
 &    + \left(1-f_\text{sig}\right)\cdot P_\text{bkg}\Big(m_\cPqt^\text{fit},m_\PW^\text{reco}\Big)\\
  = {} & f_\text{sig} \cdot \sum_{j}f_{j}P_{j}\Big(m_\cPqt^\text{fit}|m_\cPqt,\mathrm{JES}\Big)\cdot P_{j}\Big(m_\PW^\text{reco}|m_\cPqt,\mathrm{JES}\Big)\\
 & + \left(1-f_\text{sig}\right)\cdot P_\text{bkg}\Big(m_\cPqt^\text{fit}\Big)\cdot P_\text{bkg}\Big(m_\PW^\text{reco}\Big),
\end{align*}
}{
\begin{align*}
P\Big(m_\cPqt^\text{fit},m_\PW^\text{reco}|m_\cPqt,\mathrm{JES}\Big)  = {} & f_\text{sig} \cdot P_\text{sig}\Big(m_\cPqt^\text{fit},m_\PW^\text{reco}|m_\cPqt,\mathrm{JES}\Big)\\
 &    + \left(1-f_\text{sig}\right)\cdot P_\text{bkg}\Big(m_\cPqt^\text{fit},m_\PW^\text{reco}\Big)\\
  = {} & f_\text{sig} \cdot \sum_{j}f_{j}P_{j}\Big(m_\cPqt^\text{fit}|m_\cPqt,\mathrm{JES}\Big)\cdot P_{j}\Big(m_\PW^\text{reco}|m_\cPqt,\mathrm{JES}\Big)\\
 & + \left(1-f_\text{sig}\right)\cdot P_\text{bkg}\Big(m_\cPqt^\text{fit}\Big)\cdot P_\text{bkg}\Big(m_\PW^\text{reco}\Big),
\end{align*}
}
where $f_j$ with $j\in\left\{ cp,wp,un\right\}$ is the relative fraction of the three different permutation cases.
The relative fractions $f_j$ and the  probability density functions $P_{j}$ for signal are determined from simulated \ttbar events generated for nine different top-quark mass ($m_{\cPqt,\,\text{gen}}$) values and three different JES values (0.96, 1.00, and 1.04). For the probability density functions, the  $m_\cPqt^\text{fit}$ distributions are fitted with a Breit--Wigner function convolved with a Gaussian resolution function for the $cp$ case and with the sum of a Landau function and a Gaussian function with common means for the $wp$ and $un$ cases for different generated top-quark masses and jet energy scales.
The corresponding $m_\PW^\text{reco}$ distributions are distorted by the jet-selection criteria and the goodness-of-fit probability requirement and weighting because permutations with a reconstructed W-boson mass close to $80.4$\GeV are preferred by the kinematic fit.
The $m_\PW^\text{reco}$ distributions are therefore fitted with asymmetric generalized Gaussian functions.
The dependence of the parameters of the fitted functions on $m_{\cPqt,\,\text{gen}}$ and JES is then expressed in a linear function of the generated top-quark mass, JES, and the product of the two.

As the background is modeled from data, the probability density distributions for the background depend neither on the top-quark mass nor the JES.
Its $m_\cPqt^\text{fit}$ distribution is fitted by the sum of a Gamma function and a Landau function and its $m_\PW^\text{reco}$ distribution by an asymmetric Gaussian function.

In the 1D analysis, where the JES is not measured simultaneously, the top-quark mass is estimated from the minimization of $-2\ln\left\{ \mathcal{L}\left(m_\cPqt,{\mathrm{JES}=1}|\text{sample}\right)\right\}$.
In the 2D analysis the most likely top-quark mass and JES are obtained by minimizing $-2\ln\left\{ \mathcal{L}\left(m_\cPqt,\mathrm{JES}|\text{sample}\right)\right\}$. We fit a parabola (elliptic paraboloid) to extract the minimum and $1\sigma$ uncertainty from the 1D (2D) log-likelihoods.

\section{Analysis Calibration}
\label{sec:calibration}
The method is tested for possible biases and for the correct estimation of the statistical uncertainty using pseudo-experiments.
For each combination of nine different generated top-quark masses and three jet energy scales, we conduct 10\,000 pseudo-experiments using simulated \ttbar events and modeled background events from event mixing on data.
We extract $m_{\cPqt,\,\text{ext}}$ and $\mathrm{JES}_\text{ext}$ from each pseudo-experiment, which corresponds to an integrated luminosity of 3.54\fbinv. This results in 27 calibration points in the $m_{\cPqt,\, \text{gen}}$-$\mathrm{JES}$ plane.

\begin{figure}[htbp]
\begin{center}
  \includegraphics[width=0.48\textwidth]{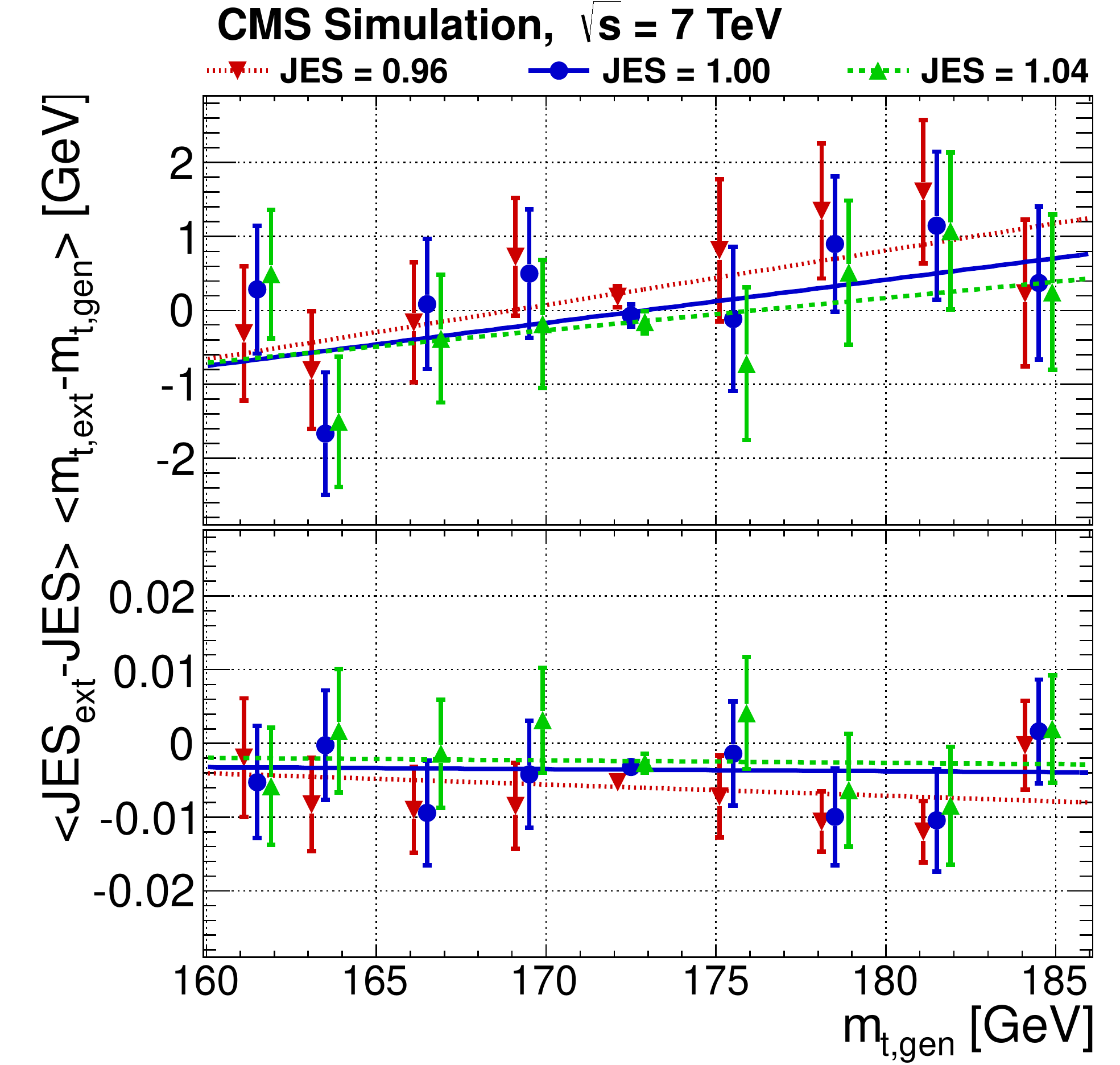}
\end{center}
\caption{\label{fig:Calibration} Difference between the extracted top-quark mass $ m_{\cPqt,\,\text{ext}}$ and the generated top-quark mass $m_{\cPqt,\,\text{gen}}$, (upper) and between the extracted and generated values of JES (lower) before calibration, for different generated top-quark masses and three different JES values. The lines correspond to linear fits which are used to correct the final likelihoods. The mass points for different JES values are shifted horizontally for clarity.
}
\end{figure}

\begin{figure}[htbp]
\includegraphics[width=0.48\textwidth]{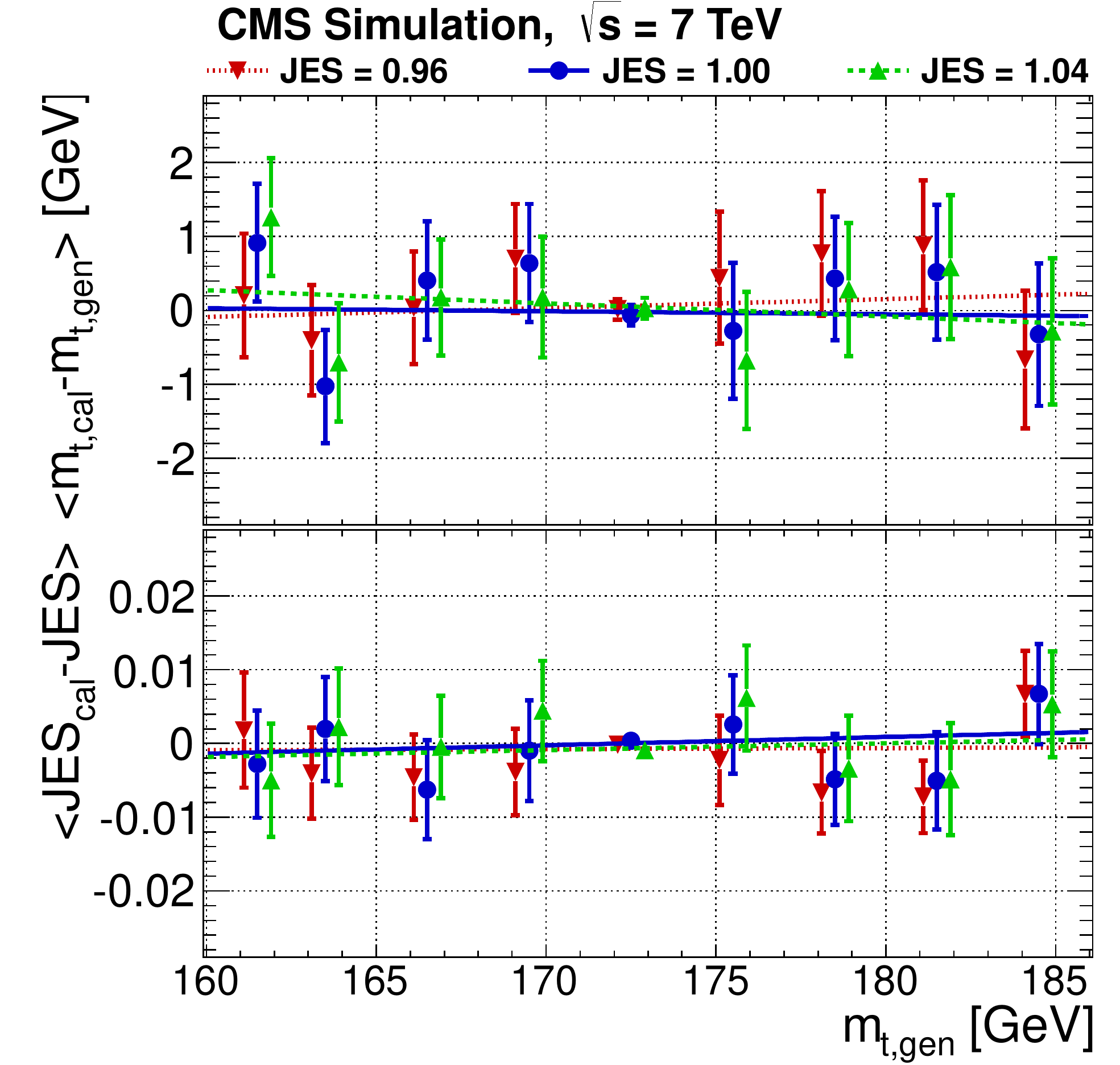}
\includegraphics[width=0.48\textwidth]{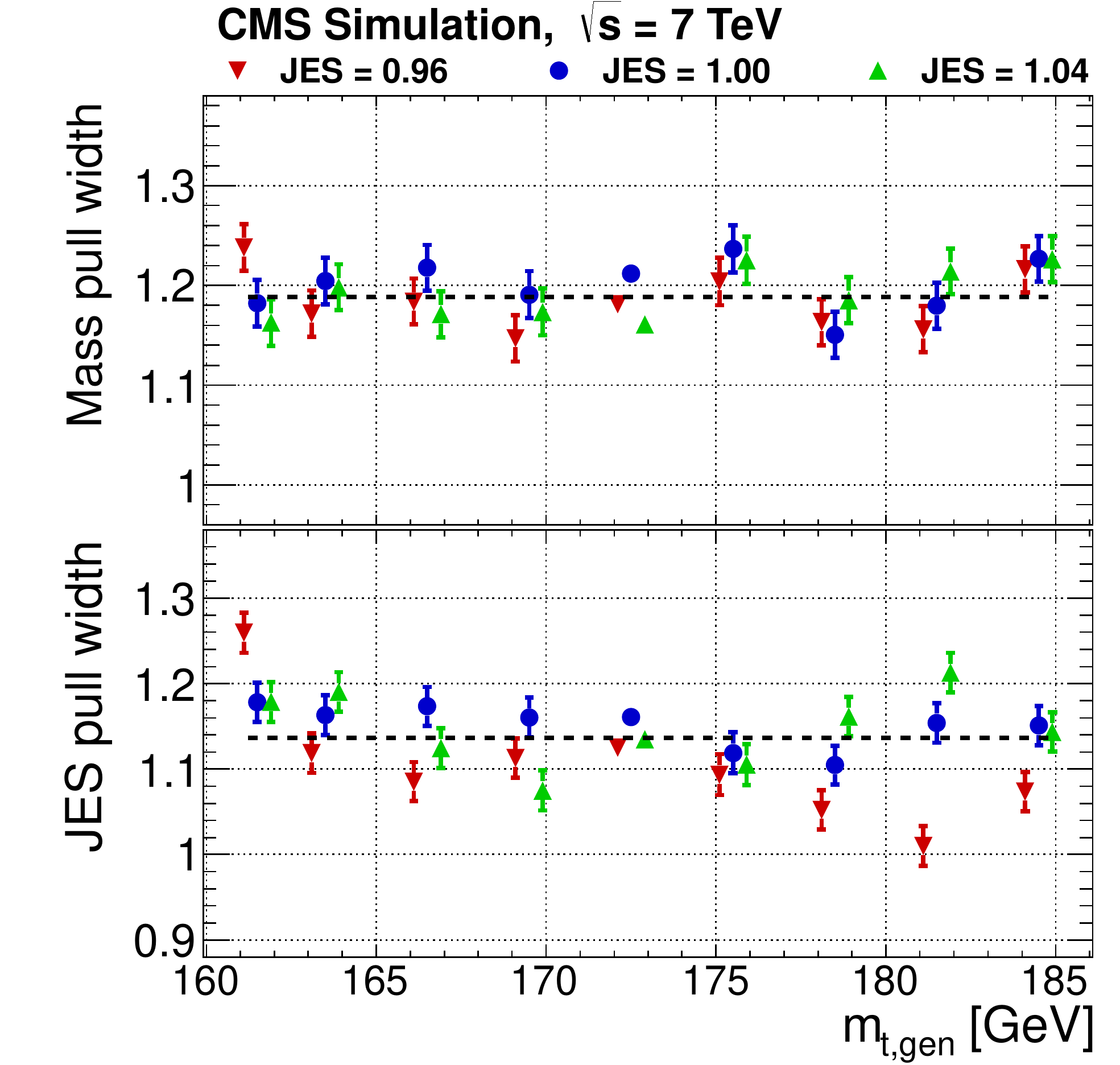}
\caption{\label{fig:Pull}  (\cmsLeft) Difference between the calibrated top-quark mass $ m_{\cPqt,\,\text{cal}}$ and the generated top-quark mass $m_{\cPqt,\,\text{gen}}$, and between the calibrated and the generated values of JES after calibration for different generated top-quark masses and three different JES values; (\cmsRight)  width of the pull distribution for the calibrated top-quark mass and for the calibrated JES for different generated top-quark masses and three different JES values. The colored lines correspond to linear fits for individual values of JES and the black line corresponds to a linear (\cmsLeft) or constant (\cmsRight) fit to all calibration points. The mass points for different JES values are shifted horizontally for clarity.
}
\end{figure}

The biases are defined as
\begin{align*}
\text{mass bias} & =   \Big<m_{\cPqt,\,\text{ext}}-m_{\cPqt,\,\text{gen}} \Big>\,;\\
\text{JES bias} & =   \Big<\mathrm{JES}_\text{ext}-\mathrm{JES} \Big>.
\end{align*}
Both mass and JES bias are plotted as a function of $m_{\cPqt,\,\text{gen}}$ for all three different JES values in Fig.~\ref{fig:Calibration}. The bias is fit with a linear function for each generated JES value. Additional small corrections for calibrating the  top-quark mass $m_{\cPqt,\,\mathrm{cal}}$ and the jet energy scale $\mathrm{JES}_\text{cal}$ are derived as linear functions of both the extracted top-quark mass and JES from these fits.
As shown in Fig.~\ref{fig:Pull} (\cmsLeft),  no further corrections are needed for the calibrated top-quark mass $m_{\cPqt,\,\text{cal}}$ and for the calibrated jet energy scale $\mathrm{JES}_\text{cal}$.

{\tolerance=800
Using pseudo-experiments with the calibrated likelihood, we fit a Gaussian function to the distribution of the pulls defined as
\begin{equation*}
\text{pull}=\frac{m_{\cPqt,\,\text{cal}}-m_{\cPqt,\,\text{gen}}}{\sigma\left(m_{\cPqt,\,\text{cal}}\right)},
\end{equation*}
 where $\sigma\left(m_{\cPqt,\,\text{cal}}\right)$ is the statistical uncertainty in an individual $m_{\cPqt,\,\text{cal}}$ for a pseudo-experiment generated at $m_{\cPqt,\,\text{gen}}$.
As depicted in Fig.~\ref{fig:Pull} (\cmsRight), we find a mass pull width of 1.19, meaning that our method underestimates the statistical uncertainty. We correct for this by dividing $-2\ln\left\{ \mathcal{L}\left(m_\cPqt,\mathrm{JES}|\text{sample}\right)\right\}$  by the square of the found mass pull width.
From these pseudo-experiments, the statistical uncertainty in the measured top-quark mass is expected to be $0.64 \pm 0.03$\GeV for the 1D analysis and $0.95 \pm 0.03$\GeV for the 2D analysis.
\par}

\section{Systematic Uncertainties}
\label{sec:systematics}
An overview of the different sources of systematic uncertainties is shown in Table~\ref{tab:Systematic-uncertainties} for the 1D analysis with a fixed JES and the 2D analysis where we estimate the top-quark mass and JES simultaneously.
The effect of a source on the efficiency to select \ttbar events and hence on the signal fraction $f_\text{sig}$ is taken into account in the evaluation.
 In general, the largest observed shifts in the top-quark mass and JES when varying the parameters studied are quoted as systematic uncertainties. If the statistical uncertainty in a shift is larger than the observed shift value we quote the statistical uncertainty in the shift instead.
The different systematic uncertainties considered as relevant for this measurement and the method to evaluate them are:
\begin{description}
\item [{{Fit calibration:}}] We propagate the statistical uncertainty of the calibration to the final measured quantities.
\item [{{Jet energy scale:}}] The effect of the uncertainty in the jet energy corrections is estimated by scaling all jet energies up and down according to their overall uncertainty~\cite{Chatrchyan:2011ds}. The scaling leads to an average JES shift of 1.2\%. We take the largest difference in measured top-quark mass as a systematic uncertainty. The systematic uncertainty in the measured JES for the 2D analysis is obtained by comparing the measured JES for the scaled samples with the expected JES shift of 1.2\%.
\item [{{b-JES:}}] The different energy responses for jets originating from light quarks (uds), bottom quarks, and gluons have been studied in simulation. It is found that the b-jet response is intermediate between the light-quark and gluon jet responses~\cite{Chatrchyan:2011ds}. Hence, the flavor uncertainty assumed for the JES determination~\cite{Chatrchyan:2011ds} to cover the transition from a gluon-dominated to a light-quark-dominated sample  also covers the transition from a sample of light quarks to one of  bottom quarks. Thus, the energies of all b jets are scaled up and down by this flavor uncertainty in simulation that ranges from 0.2\% to 1.2\%.
\item [{{Jet energy resolution:}}] The jet energy resolution in simulation is degraded by 7\% to 20\% depending on $\eta$ to match the resolutions found in \cite{Chatrchyan:2011ds}. To account for the resolution uncertainty, two additional shifts corresponding to ${\pm}1\sigma$ are evaluated.
\item [{{b tagging:}}] The threshold on the CSVT tagger is varied in order to reflect an uncertainty of the b-tag efficiency of 3\%~\cite{Chatrchyan:2012jua}.
\item [{{Trigger:}}] The uncertainty in the turn-on of the jet triggers in data is estimated by raising the jet \pt cuts on the 4th, 5th, and 6th jets separately by 2\GeV in the \ttbar simulation. Each increase lowers the selection efficiency by 7 to 10\% covering the uncertainty of 5\% found in a dedicated study for the \ttbar cross section measurement in this channel~\cite{Chatrchyan:2013ual}. We quote the quadratic sum of the observed shifts in top-quark mass and JES from each increase as systematic uncertainty.
\item [{{Pileup:}}] To estimate the uncertainties associated with the determination of the number of pileup events and with the weighting procedure, the average number of expected pileup events (8.1) is varied by ${\pm}5$\%.
\item [{{Parton distribution functions:}}] The simulated events have been generated using the CTEQ 6.6L parton distribution functions (PDFs)~\cite{Nadolsky:2008zw}.
The uncertainty in this PDF set is described by up/down variation of 22 orthogonal parameters resulting in 22 pairs of additional PDFs.
The events are weighted for agreement with the additional PDFs and half of the difference in top-quark mass and JES of each pair is quoted as systematic uncertainties. The systematic uncertainties stemming from each pair are added in quadrature.
\item [{{Renormalization and factorization scale:}}] The dependence of the result on the renormalization and factorization scale used in the \ttbar simulation is studied by varying the scale choice for the hard scattering and for parton showering by a factor 0.5 and 2.0. The variation of these parameters in simulation reflects also the uncertainty in the amount of initial state and final state radiation.
\item [{{ME-PS matching threshold:}}] In the \ttbar simulation, the matching threshold used for interfacing the matrix elements generated with \MADGRAPH and the \PYTHIA parton showering is varied by factors of 0.5 and 2.0 compared to the default threshold.
\item [{{Underlying event:}}] Non-perturbative QCD effects are taken into account by tuning \PYTHIA to measurements of the underlying event~\cite{Chatrchyan:2011id}. The uncertainties are estimated by comparing in simulation two tunes with increased and decreased underlying event activities to a central tune (the Perugia 2011 tune to the Perugia 2011 mpiHi and Perugia 2011 Tevatron tunes~\cite{Skands:2010ak}).
\item [{{Color reconnection effects:}}] {\tolerance=800 The uncertainties that arise from different modeling of color reconnection effects~\cite{Skands:2007zg} are estimated by comparing in simulation an underlying event tune with color reconnection to a tune without it (the Perugia 2011 and Perugia 2011NoCR tunes~\cite{Skands:2010ak}).\par}
\item [{{Multijet background:}}] After the final selection, a signal fraction of 54\% is expected from simulation. The signal fraction is varied between 49\% and 59\%, corresponding to the uncertainties of the theoretical predictions of the \ttbar cross section, the value of the top-quark mass, and the luminosity.  In addition, we study the effect of \ttbar events in the input sample used for the event mixing. To estimate the effect, the event mixing is performed in simulation on a \ttbar sample and alternative probability density distributions are derived from this sample for the background. This variation also accounts for the small shape differences observed for the event mixing technique on the additional \bbbar sample.
\end{description}

\begin{table*}[bthp]
\topcaption{\label{tab:Systematic-uncertainties}Overview of systematic uncertainties. The total is defined by adding in quadrature the contributions from all sources, by choosing for each the larger of the estimated shift or its statistical uncertainty, as indicated by the bold script.}
\centering
\begin{tabular}{l|c|cc}
  & 1D analysis         & \multicolumn{2}{c}{2D analysis} \\
\cline{2-4}
 & $\delta_{m_\cPqt}$ (\GeV) & $\delta_{m_\cPqt}$ (\GeV) & $\delta_\mathrm{JES}$ \\
\hline
\hline
Fit calibration                         & $\mathbf{0.13}         $ & $\mathbf{0.14}         $ & $\mathbf{0.001}$  \\
Jet energy scale                        & $\mathbf{0.97} \pm 0.06$ & $0.09 \pm \mathbf{0.10}$ & $\mathbf{0.002} \pm 0.001 $\\
b-JES                                   & $\mathbf{0.49} \pm 0.06$ & $\mathbf{0.52} \pm 0.10$ & $\mathbf{0.001} \pm 0.001 $ \\
Jet energy resolution                   & $\mathbf{0.15} \pm 0.06$ & $\mathbf{0.13} \pm 0.10$ & $\mathbf{0.003} \pm 0.001 $\\
b tagging                               & $0.05 \pm \mathbf{0.06}$ & $0.04 \pm \mathbf{0.10}$ & $\mathbf{0.001} \pm 0.001 $\\
Trigger                                 & $\mathbf{0.24} \pm 0.06$ & $\mathbf{0.26} \pm 0.10$ & $\mathbf{0.006} \pm 0.001 $\\
Pileup                                  & $0.05 \pm \mathbf{0.06}$ & $0.09 \pm \mathbf{0.10}$ & $\mathbf{0.001} \pm 0.001 $\\
Parton distribution functions           & $0.03 \pm \mathbf{0.06}$ & $0.07 \pm \mathbf{0.10}$ & $\mathbf{0.001} \pm 0.001 $\\
Renormalization and factorization scale & $0.08 \pm \mathbf{0.22}$ & $0.31 \pm \mathbf{0.34}$ & $\mathbf{0.005} \pm 0.003 $\\
ME-PS matching threshold                & $\mathbf{0.24} \pm 0.22$ & $0.29 \pm \mathbf{0.34}$ & $0.001 \pm \mathbf{0.003} $\\
Underlying event                        & $\mathbf{0.20} \pm 0.12$ & $\mathbf{0.42} \pm 0.20$ & $\mathbf{0.004} \pm 0.002 $\\
Color reconnection effects              & $0.04 \pm \mathbf{0.15}$ & $\mathbf{0.58} \pm 0.25$ & $\mathbf{0.006} \pm 0.002 $\\
Multijet background                     & $\mathbf{0.13} \pm 0.06$ & $\mathbf{0.60} \pm 0.10$ & $\mathbf{0.006} \pm 0.001 $\\
\hline
\hline
Total                                   & 1.21 & 1.23 & 0.013 \tabularnewline
\end{tabular}
\end{table*}

As expected, the main systematic uncertainty in the 1D analysis stems from the uncertainty in the jet energy scale and the 2D analysis reduces this uncertainty to a small \PT- and $\eta$-dependent JES uncertainty, but leads to a larger statistical uncertainty in the measured top-quark mass.
Within the statistical precision of the uncertainty evaluation, most other systematic uncertainties are compatible. The variation of the signal fraction $f_{\text{sig}}$ contributes 0.11\GeV (0.10\GeV) to the systematic uncertainty on the multijet background in the 1D (2D) analysis justifying that $f_{\text{sig}}$ is kept fixed in the likelihood method.
However, the  2D analysis has increased uncertainties for color reconnection effects and the shape of the multijet background.
Due to the W-boson mass constraint in the kinematic fit, only the color reconnection effects for the b quarks affect the 1D analysis.
For the 2D analysis, the JES estimation from the reconstructed W-boson masses results in an additional dependence on color reconnection effects for the light quarks and, hence, an increased systematic uncertainty. Similarly, the additional uncertainty in the modeling of the distribution of the reconstructed W-boson masses for the background gets propagated into the measured top-quark mass for the multijet background uncertainty.

Overall, the systematic uncertainties for both methods are very similar in size.
This is in contrast to the CMS measurement in the lepton+jets channel~\cite{Chatrchyan:2012cz} where the simultaneous fit of the top-quark mass and the JES leads to a reduction of the systematic uncertainty by 40\%.
However, the  jets are required to have a higher minimum transverse momentum in the all-jets channel, which leads to a reduced uncertainty in the JES in the 1D analysis compared to the previous work~\cite{Chatrchyan:2012cz}.
In addition, the tighter jet criteria in the all-jets measurement have a stronger impact on the $m_\PW^\text{reco}$ distribution, making the JES estimation more sensitive to changes in the simulation.

\section{Results}
\label{sec:results}
\begin{figure*}[thbp]
\centering
\includegraphics[width=0.48\textwidth]{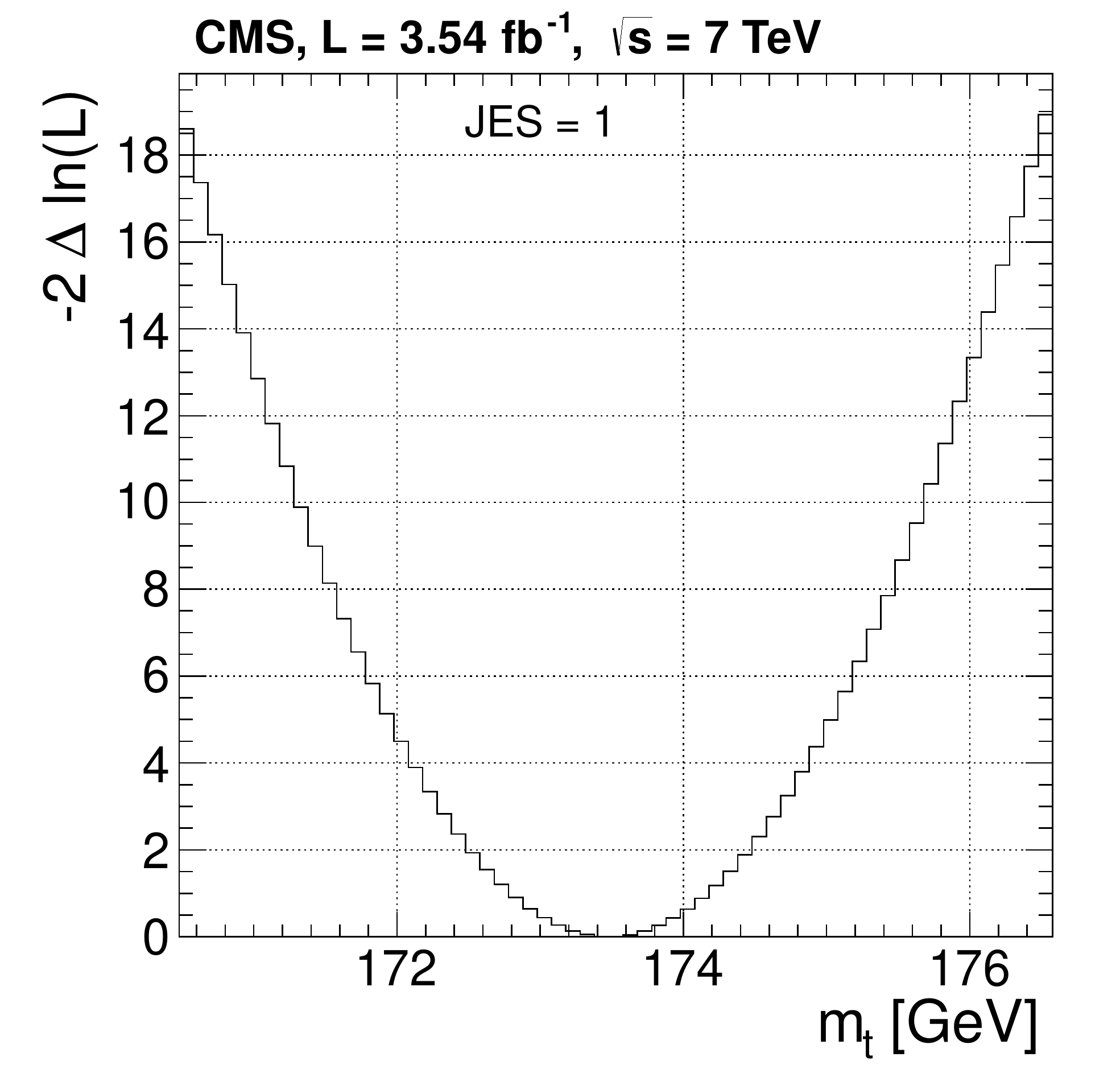}
\includegraphics[width=0.48\textwidth]{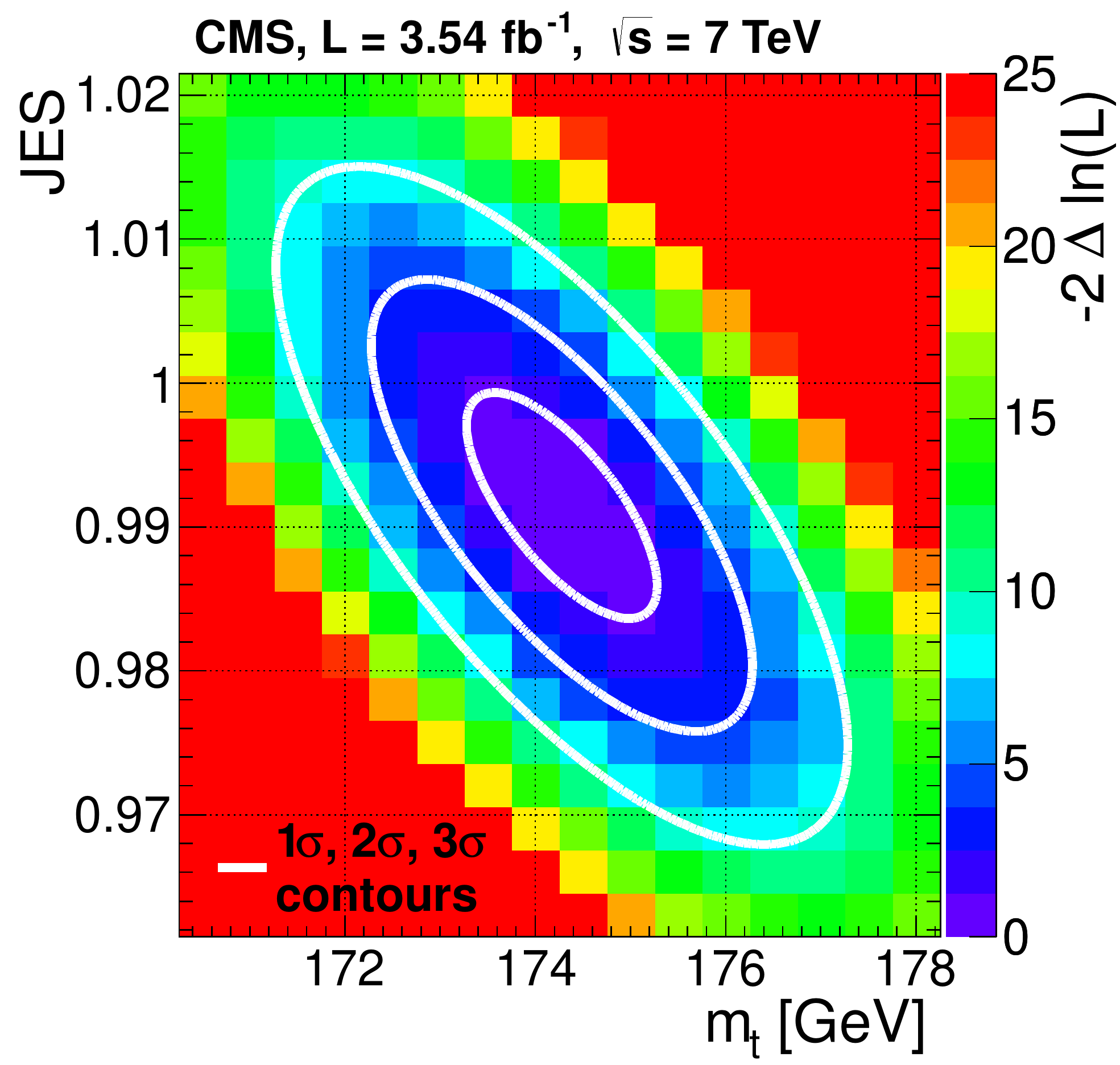}
\caption{\label{fig:result} (left) The 1D likelihood profile with the JES fixed to unity and (right) the 2D likelihood.
The contours correspond to $1\sigma$, $2\sigma$, and $3\sigma$ statistical uncertainties.}
\end{figure*}

From  the selected 2418 events we measure with the jet energy scale fixed to the nominal value of JES $= 1$:
\begin{equation*}
m_\cPqt  =  173.49 \pm 0.69\stat\pm 1.21\syst\GeV
\end{equation*}
The overall uncertainty of the presented 1D analysis is 1.39\GeV. The likelihood profile used in the 1D analysis is shown in Fig.~\ref{fig:result} (left).

A simultaneous fit of the top-quark mass and JES to the same data yields:
\begin{align*}
m_\cPqt       & =  174.28 \pm 1.00\,\text{(stat.+JES)} \pm 1.23\syst\GeV\\
\mathrm{JES} & =   0.991 \pm 0.008\stat \pm 0.013\syst.
\end{align*}
The measured JES confirms the JES for particle-flow jets in data measured in events where a Z boson or photon is produced together with one jet~\cite{Chatrchyan:2011ds}. In the 2D analysis the overall uncertainty in the top-quark mass is 1.58\GeV. As the top-quark mass and JES are measured simultaneously, the uncertainty in the top-quark mass combines the statistical uncertainties arising from both components. Figure~\ref{fig:result} (right) shows the 2D likelihood obtained from data. The measured top-quark masses in both analyses are in agreement, but the 1D analysis has a better precision than the  2D analysis.

We use the Best Linear Unbiased Estimate technique~\cite{Lyons:1988rp} to combine the 1D result presented in this paper with the CMS measurements in the dilepton channel based on \text{2010}~\cite{Chatrchyan:2011nb} and \text{2011}~\cite{Chatrchyan:2012ea} data, and the measurement in the lepton+jets channel~\cite{Chatrchyan:2012cz}.
Most of the systematic uncertainties listed in Table~\ref{tab:Systematic-uncertainties} are assumed to be fully correlated among the three input measurements.
Exceptions are the uncertainties in pileup, for which we assign full correlation between the 2011 analyses but no correlation with the 2010 analysis, since the pileup conditions and their treatments differ.
In addition, the statistical uncertainty in the in situ fit for the JES and the uncertainties in the mass calibration, the background normalization from control samples in data in the dilepton, and the background prediction in the all-jets analysis are treated as uncorrelated systematic uncertainties.
The combination of the four measurements yields a mass of $m_\cPqt = 173.54 \pm 0.33\stat\pm 0.96\syst$\GeV. It has a $\chi^2$ of 1.4 for three degrees of freedom, which corresponds to a probability of 71\%.
\begin{figure*}[htbp]
\centering
\includegraphics[width=0.8\textwidth]{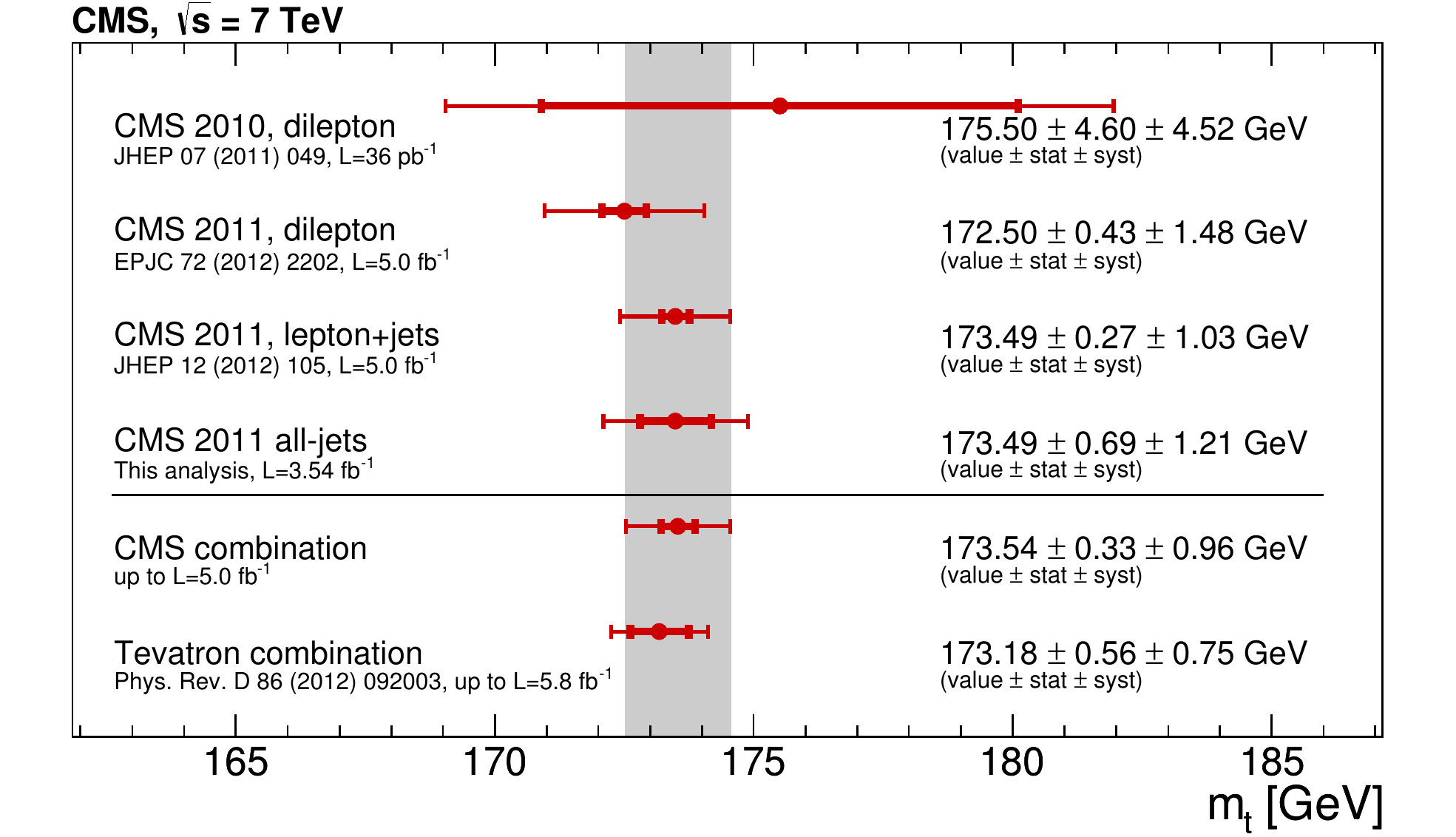}
\caption{\label{fig:combination} Overview of the CMS top-quark mass measurements, their combination that is also shown as the shaded band, and the Tevatron average. The inner error bars indicate the statistical uncertainty, the outer error bars indicate the total uncertainty. The statistical uncertainty in the in situ fit for the JES is treated as a systematic uncertainty.}
\end{figure*}
Figure~\ref{fig:combination} gives an overview of the input measurements and the combined result.

\section{Summary}
A measurement of the top-quark mass is presented using events with at least six jets in the final state, collected by CMS in $\Pp\Pp$ collisions at $\sqrt{s} = 7$\TeV in 2011.
The complete kinematic properties of each event are reconstructed using a constrained fit to a \ttbar hypothesis.
For each selected event a likelihood is calculated as a function of assumed values of the top-quark mass.
From a data sample corresponding to an integrated luminosity of 3.54\fbinv, 2418 candidate events are observed and the mass of the top quark is measured to be $m_\cPqt = 173.49 \pm 0.69\stat \pm 1.21\syst\GeV.$
This result for $m_\cPqt$ is consistent with the Tevatron average~\cite{PhysRevD.86.092003}, with the ATLAS measurement in the lepton+jets channel~\cite{ATLAS:2012aj}, and with CMS measurements in the lepton+jets~\cite{Chatrchyan:2012cz} and dilepton~\cite{Chatrchyan:2012ea} channels. To date, this measurement constitutes the most precise determination of the top-quark mass in the all-jets channel. A combination with the three previously published CMS measurements~\cite{Chatrchyan:2011nb,Chatrchyan:2012ea,Chatrchyan:2012cz} yields a mass of $m_\cPqt = 173.54 \pm 0.33 \stat\pm 0.96\syst = 173.54 \pm 1.02$\GeV, consistent with the Tevatron average~\cite{PhysRevD.86.092003} and with similar precision.

\section*{Acknowledgements}
{\tolerance=800
We congratulate our colleagues in the CERN accelerator departments for the excellent performance of the LHC and thank the technical and administrative staffs at CERN and at other CMS institutes for their contributions to the success of the CMS effort. In addition, we gratefully acknowledge the computing centres and personnel of the Worldwide LHC Computing Grid for delivering so effectively the computing infrastructure essential to our analyses. Finally, we acknowledge the enduring support for the construction and operation of the LHC and the CMS detector provided by the following funding agencies: BMWF and FWF (Austria); FNRS and FWO (Belgium); CNPq, CAPES, FAPERJ, and FAPESP (Brazil); MES (Bulgaria); CERN; CAS, MoST, and NSFC (China); COLCIENCIAS (Colombia); MSES (Croatia); RPF (Cyprus); MoER, SF0690030s09 and ERDF (Estonia); Academy of Finland, MEC, and HIP (Finland); CEA and CNRS/IN2P3 (France); BMBF, DFG, and HGF (Germany); GSRT (Greece); OTKA and NKTH (Hungary); DAE and DST (India); IPM (Iran); SFI (Ireland); INFN (Italy); NRF and WCU (Republic of Korea); LAS (Lithuania); CINVESTAV, CONACYT, SEP, and UASLP-FAI (Mexico); MSI (New Zealand); PAEC (Pakistan); MSHE and NSC (Poland); FCT (Portugal); JINR (Dubna); MON, RosAtom, RAS and RFBR (Russia); MESTD (Serbia); SEIDI and CPAN (Spain); Swiss Funding Agencies (Switzerland); NSC (Taipei); ThEPCenter, IPST, STAR and NSTDA (Thailand); TUBITAK and TAEK (Turkey); NASU (Ukraine); STFC (United Kingdom); DOE and NSF (USA).

Individuals have received support from the Marie-Curie programme and the European Research Council and EPLANET (European Union); the Leventis Foundation; the A. P. Sloan Foundation; the Alexander von Humboldt Foundation; the Belgian Federal Science Policy Office; the Fonds pour la Formation \`a la Recherche dans l'Industrie et dans l'Agriculture (FRIA-Belgium); the Agentschap voor Innovatie door Wetenschap en Technologie (IWT-Belgium); the Ministry of Education, Youth and Sports (MEYS) of Czech Republic; the Council of Science and Industrial Research, India; the Compagnia di San Paolo (Torino); the HOMING PLUS programme of Foundation for Polish Science, cofinanced by EU, Regional Development Fund; and the Thalis and Aristeia programmes cofinanced by EU-ESF and the Greek NSRF.
\par}
\bibliography{auto_generated}   

\cleardoublepage \appendix\section{The CMS Collaboration \label{app:collab}}\begin{sloppypar}\hyphenpenalty=5000\widowpenalty=500\clubpenalty=5000\input{TOP-11-017-authorlist.tex}\end{sloppypar}
\end{document}

%% file: TOP-11-017-authorlist.tex
\textbf{Yerevan Physics Institute,  Yerevan,  Armenia}\\*[0pt]
S.~Chatrchyan, V.~Khachatryan, A.M.~Sirunyan, A.~Tumasyan
\vskip\cmsinstskip
\textbf{Institut f\"{u}r Hochenergiephysik der OeAW,  Wien,  Austria}\\*[0pt]
W.~Adam, T.~Bergauer, M.~Dragicevic, J.~Er\"{o}, C.~Fabjan\cmsAuthorMark{1}, M.~Friedl, R.~Fr\"{u}hwirth\cmsAuthorMark{1}, V.M.~Ghete, N.~H\"{o}rmann, J.~Hrubec, M.~Jeitler\cmsAuthorMark{1}, W.~Kiesenhofer, V.~Kn\"{u}nz, M.~Krammer\cmsAuthorMark{1}, I.~Kr\"{a}tschmer, D.~Liko, I.~Mikulec, D.~Rabady\cmsAuthorMark{2}, B.~Rahbaran, C.~Rohringer, H.~Rohringer, R.~Sch\"{o}fbeck, J.~Strauss, A.~Taurok, W.~Treberer-Treberspurg, W.~Waltenberger, C.-E.~Wulz\cmsAuthorMark{1}
% ...
\vskip\cmsinstskip
% ...
1:~~Also at Vienna University of Technology, Vienna, Austria\\
2:~~Also at CERN, European Organization for Nuclear Research, Geneva, Switzerland\\
% ...
